\documentclass{article}

\usepackage{arxiv}

\usepackage[utf8]{inputenc} 
\usepackage[T1]{fontenc}    
\usepackage{hyperref}       
\usepackage{url}            
\usepackage{booktabs}       
\usepackage{amsfonts}       
\usepackage{nicefrac}       
\usepackage{microtype}      
\usepackage{lipsum}
\usepackage{amsmath}
\usepackage{graphicx}
\usepackage{epstopdf}
\usepackage{cite}

\title{Hybrid modes in Al nanoparticles arrays}

\author{A.~E.~Ershov\\
Institute of Computational Modeling SB RAS, Krasnoyarsk 660036, Russia\\
Siberian Federal University, Krasnoyarsk, 660041, Russia\\
\And
V.~S.~Gerasimov\\
Institute of Computational Modeling SB RAS, Krasnoyarsk 660036, Russia\\
Siberian Federal University, Krasnoyarsk, 660041, Russia\\
\And
R.~G.~Bikbaev\\
L.V. Kirensky Institute of Physics, Federal Research Center KSC SB RAS, 660036, Krasnoyarsk, Russia
\\Siberian Federal University, Krasnoyarsk, 660041, Russia
\And
S.~P.~Polyutov\\
L.V. Kirensky Institute of Physics, Federal Research Center KSC SB RAS, 660036, Krasnoyarsk, Russia
\\Siberian Federal University, Krasnoyarsk, 660041, Russia
\And
S.~V.~Karpov\\
L.V. Kirensky Institute of Physics, Federal Research Center KSC SB RAS, 660036, Krasnoyarsk, Russia
\\Siberian Federal University, Krasnoyarsk, 660041, Russia
\\Siberian State University of Science and Technology, 660014, Krasnoyarsk, Russia
}

\begin{document}
\maketitle

\begin{abstract}
The mechanisms of coupling between the lattice modes of a two-dimensional (2D) array consisting of Al nanoparticles and the localized modes of individual Al nanoparticles have been studied in detail. 
The results have been obtained employing the finite-difference time-domain method (FDTD) and the generalized Mie theory.
It was shown that interactions of single particles with 2D lattice modes significantly change the  extinction spectra  depending on the particle radius and the lattice period. 
The Rayleigh anomalies of higher orders contribute to  formation of hybrid modes resulting in increase of the extinction efficiency in short wavelength  range of the spectrum. 
It is shown that high intensity magnetic modes are excited in aluminum nanoparticles arrays.
The patterns of spatial electromagnetic field distribution at the frequencies of hybrid modes have been studied.
We note that comprehensive understanding the mode coupling mechanisms in  arrays paves the way for engineering different types of modern photonic devices with controllable optical properties. 
\end{abstract}


\section{Introduction}

It is generally known that appearance of electromagnetic field modes of different orders is possible in a course of interaction of external optical radiation with single isolated nanoparticles. For spherical particles the first-order modes are electro- and magnetodipole ones~\cite{Evlyukhin2012}. 
The excitation of higher order modes can be achieved by changing the particles size and the shape of the incident wave front. 
In the case of an ordered structure (e.g., the lattice) the interaction of the local modes of each individual particle with the collective modes of the entire structure can lead to the emergence of the collective resonances first described by Rayleigh~\cite{Rayleigh1907} and Wood~\cite{Wood1902}.
The appearance of these resonances in periodic arrays of plasmonic nanoparticles has been predicted in Refs~\cite{Zou2004,Zou2004a,Markel2005} and  then shown experimentally in Refs.~\cite{Auguie2008,Kravets2008,Chu2008}. 
Later these lattice resonances were studied in a wide range of periodic nanostructures with different types of unit cells: single~\cite{Khlopin2017} or paired (in a stack)~\cite{Zhang2011} nanodisks, the cylinders with the core-shell structure~\cite{Lin2015}, dimers~\cite{Humphrey2016,Mahi2017}, and more complicated configurations~\cite{Grigoriev2013,Wang2015d,Nicolas2015,Guo2017,Gerasimov19JQSRT}. 

Such the structures have found wide applications in a number of areas, for example, in IR spectroscopy~\cite{Adato2009}, narrowband light absorption~\cite{Li2014}, sensorics~\cite{Thackray2014,Gutha2017}, lasers~\cite{Zhou2013} and enhanced fluorescence~\cite{Vecchi2009,Laux2017}, just to name a few.
Notoriously that the conventional plasmonic materials are silver and gold, but in these latter days the alternative plasmonic materials such as transparent conductive oxides (AZO, GZO, ITO)~\cite{Lin2016a} and titanium nitride  which allows to obtain high-Q modes ($Q= 10^4$) in the telecommunication band~\cite{Zakomirnyi17APL} are of great interest. 
Aluminum plasmonics emerged quite recently~\cite{Knight2014,Gerard2015}.
Interest to this material is due to the fact that the plasma frequency of aluminum is higher than a one of silver or gold, allowing to observe the plasmon resonance in the ultraviolet (UV) region of the spectrum. 
This feature can be used, for example,  in  photocatalysis and for studying of organic and biological systems that exhibit strong UV absorption~\cite{Geddes2002,LAKOWICZ2005171,Fort_2007}. 
The interest to aluminum in that respect is also due to its relative cheapness and feasibility that opens wide  opportunities for manufacturing and mass production in such promising areas as color printing~\cite{Tan2014,Mahani2018,Song2019}, photovoltaics \cite{Kochergin2011,Uhrenfeldt2015,Zhang2015b,Zhan2016}, thermoplasmonics~\cite{Wiecha2017}, holography~\cite{Zhang2019}. 
 The plasmonic properties of Al structures were extensively studied in numerous articles: single nanoparticles with various shapes~ \cite{Ekinci2008,Langhammer2008,Maidecchi2013,Knight2014,Campos2017,Clark2018,Pathak2019,Clark2019}, dimers~\cite{Ross2014a}, heterodimers~\cite{Flauraud2017}, and arrays~\cite{Kannegulla2017,Khlopin2017}.
Surface lattice resonances (SLRs) with active tuning~\cite{Yang2016,Tseng2017} has been shown to cover wide range of frequencies in Al and used for nonlinear optics~\cite{Huttunen2019a}, lasers~\cite{Knudson2019,Li2019a,Li2019b}, temperature sensing~\cite{Murai2019}, photoluminescence~\cite{Lozano2013,Kawachiya2018,Kawachiya2019,Zhang2019}, light emission~\cite{Rodriguez2014} and confinement~\cite{Murai2019a}, quantum electrodynamics~\cite{Todisco2016}, sensor on Al metal film holes~\cite{Fu2019}.
The multipoles in single Al nanoparticle are studied in Ref.~\cite{Martin2014}. The methods for synthesis can be found in Ref.~\cite{Martin2013}, while the methods to  decrease the losses in Al nanoparticles are described in Ref.~\cite{Zhang2017}.

It is evident that practical implementation and design of devices based on periodic arrays requires the understanding both of how the modes of individual particles in the array interact with the lattice modes and how to control the hybrid modes. 

The goal of our paper is the studying  of interaction of local modes (both electric and magnetic) of individual particles with the collective modes of the periodic lattice structure.
The optical properties of hybrid modes for different radii of the particles and lattice period have been scrutinized. The results were obtained by analytically accurate generalized Mie theory and commercial FDTD package (Finite-Difference Time-Domain Method). 
This allows to determine contribution of the each partial hybrid spherical mode to the extinction spectra and to engineer the extinction spectra of the structure. 

\section{Computational model}

\subsection{The structure}

\begin{figure}[h]
\centerline{
\includegraphics[width=80mm]{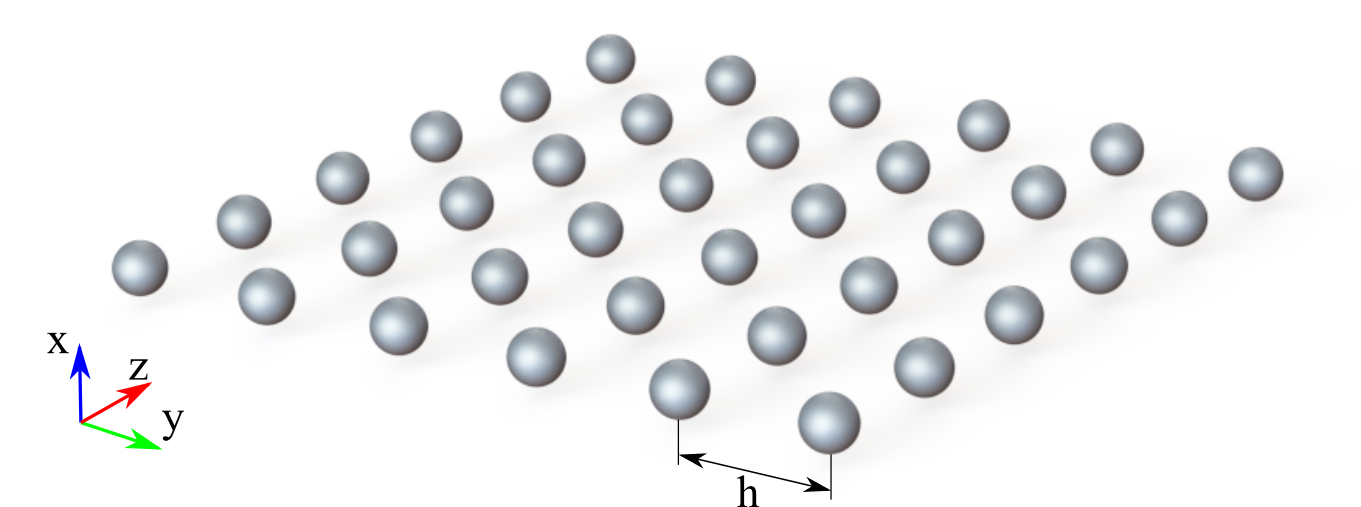}}
\caption{Sketch view of the square Al particles array.}
\label{fig:Figure1}
\end{figure}

Schematic representation of the structure used in our simulations is shown in Fig.~\ref{fig:Figure1}:  regular square 2D array of aluminum nanospheres with radius $R$ and period $h$ along $x$ and $y$ axis, respectively, is immersed in the dielectric media with permittivity $\varepsilon_{m} = 2.25$ (optical glass) and we used in simulations tabulated values for $\varepsilon$ of Al~\cite{Smith1997TheAluminum}.

\subsection{Definition of the extinction}

In this paper we consider the extinction spectra of both single nanoparticles and periodic lattices (finite and infinite) using generalized Mie theory and Finite-Difference Time-Domain (FDTD) method.

Here are the different conceptions for simulations of scattering and transmission cross-sections for finite and infinite periodic structures (and therefore extinction) which do not coincide with each other. 
In the case of infinite lattices the following definition of extinction, scattering, and transmission are commonly used.
The transmission cross-section is determined by the power of the electromagnetic field (related to the intensity of the incident radiation) that pass through the periodic structure. And the scattering cross-section~--- the remaining power of the electromagnetic field.

In the case of finite structures another definitions are used. 
The electromagnetic field outside the particles consists of incident and scattered ones:

\begin{equation}
\begin{split}
    {\bf E} = {\bf E}_i + {\bf E}_{sca},\\
    {\bf H} = {\bf H}_i + {\bf H}_{sca}.
\end{split}
\label{eq2}
\end{equation}
Here ${\bf E}_i$, ${\bf H}_i$ is the incident electromagnetic field which is defined as the electromagnetic field in the interparticle medium in the absence of the particles.
In this case the extinction cross-section is determined by the power of the field (${\bf E}_{sca}$, ${\bf H}_{sca}$).

To compare the spectra of finite structures (both single particle and finite array) with the spectra of infinite structures it is necessary to choose one of the approaches described above that will be applied to both types of structures.

The first definition is not generally applicable to finite structures without introducing additional conventions (for example, introducing a structure plane for a single particle).
The second definition can be applied to both~--- infinite and finite structures. Therefore, it was chosen as the definition of extinction in all our calculations. However, for infinite structures, this method has an important feature which should be taken into account: the coefficient of extinction can exceed an unity, nevertheless, it should not be perceived as an error.
Thus, the amplitude of the full field passed through the periodic structure will not go beyond the amplitude of the incident field, but may have a different phase. 
As a result, amplitude of the scattered field 
introduced in Eq.~(\ref{eq2}) may exceed the amplitude of the incident field due to interference effect. 

\begin{figure}
\centering
\includegraphics[width=120mm]{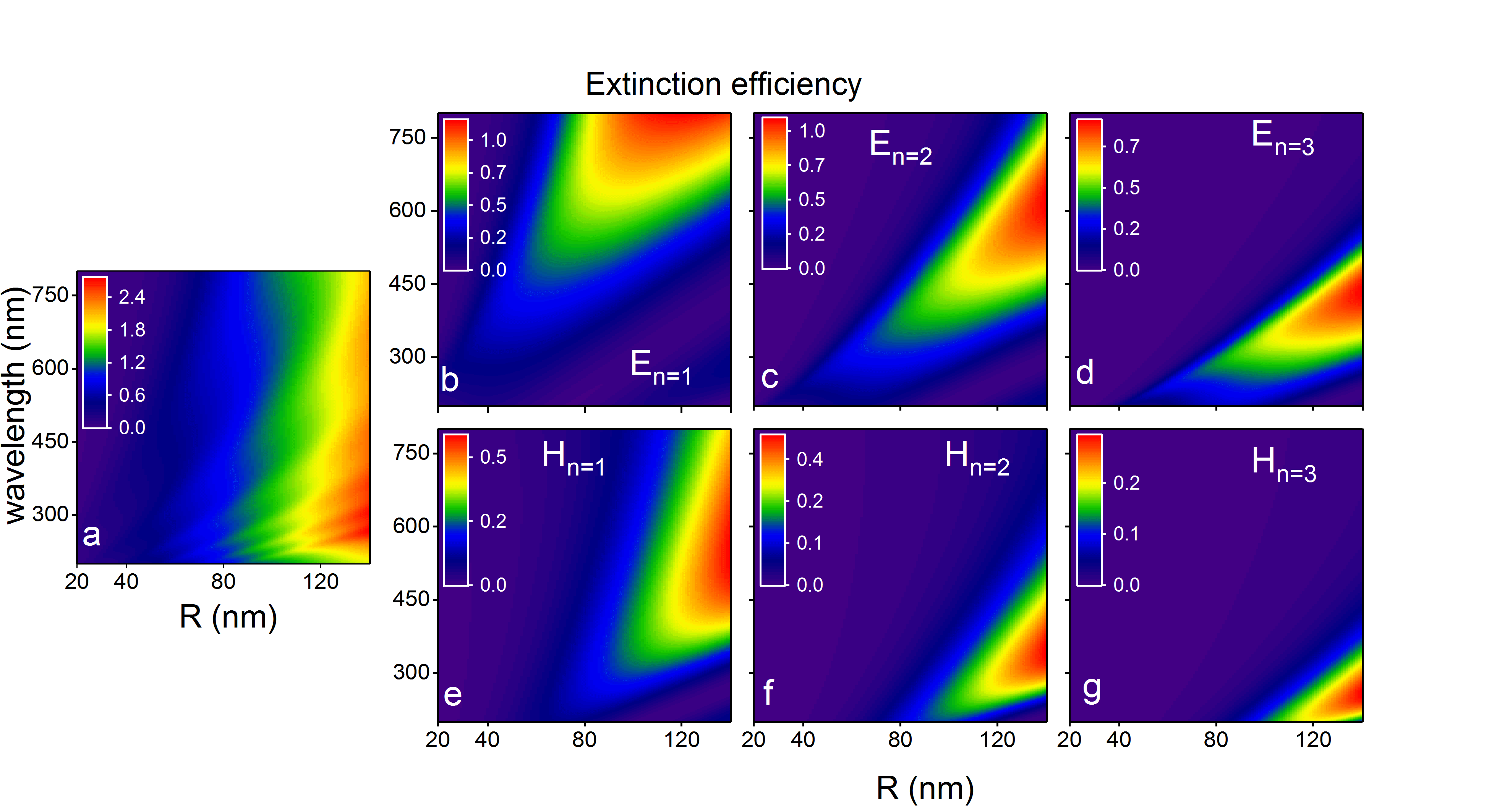}
\caption{Extinction efficiency decomposed in a set of spherical multipoles for single aluminum nanospheres of different radius $R$ and wavelength of incident light.}
\label{fig:single_mult}
\end{figure}

\begin{figure}
\centerline{
\includegraphics[width=80mm]{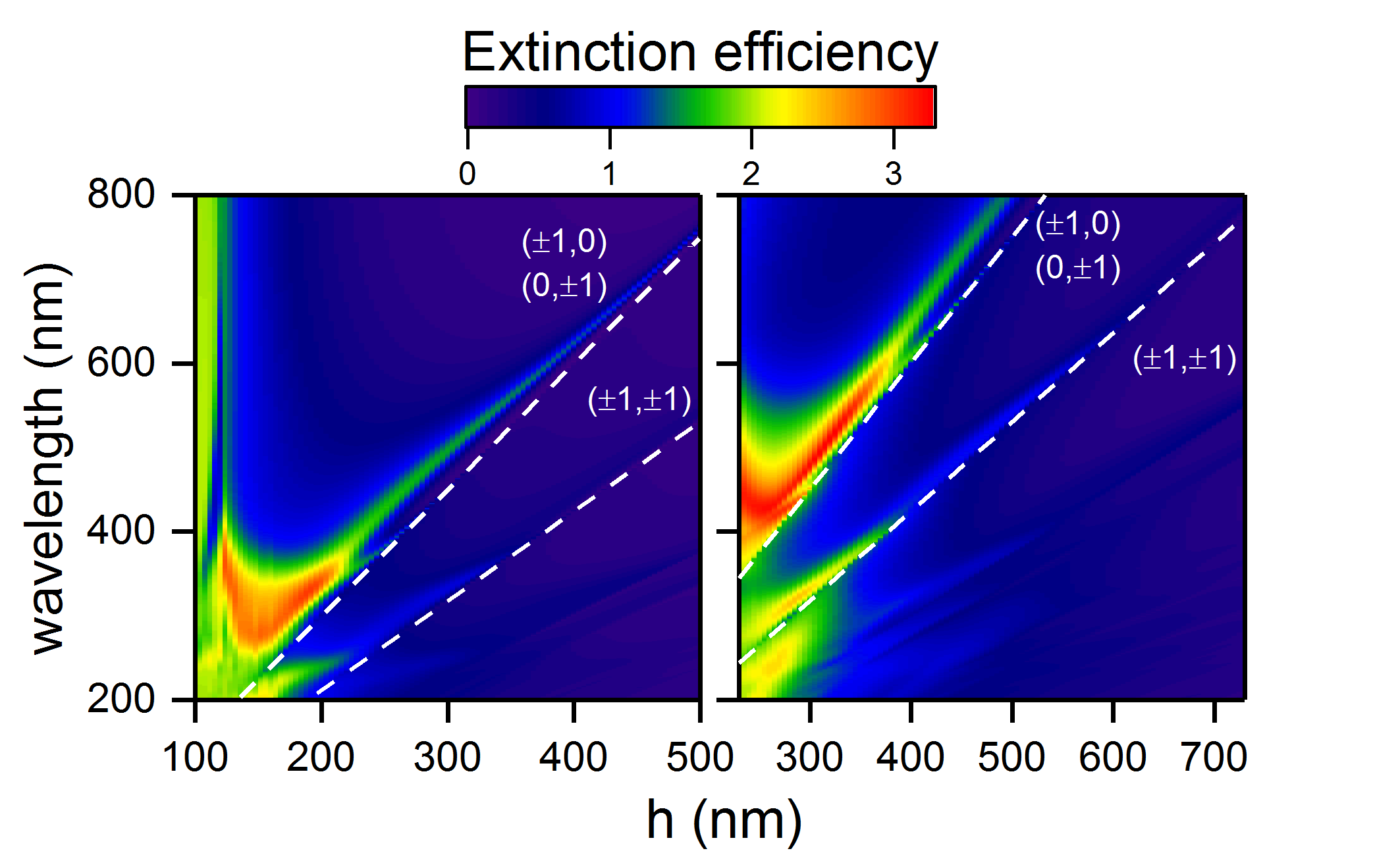}}
\caption{Extinction efficiency of Al NP's array with different period $h$. NP's radius  $R=60$~nm (left) and $R=110$~nm (right), respectively.}
\label{fig:ot_p}
\end{figure}

\subsection{Finite-Difference Time-Domain Method}

To calculate extinction spectra of infinite nanosphere arrays commercial FDTD package was used~\cite{lumerical}.
Total field scattered field (TFSF) light source (see~\cite{Potter2017}) was applied to separate (${\bf E}_i$, ${\bf H}_i$) and (${\bf E}_{sca}$, ${\bf H}_{sca}$) in Eq.~\eqref{eq2} and to find $\sigma_{sca}$ and $\sigma_{ext}$. 

The array illuminated from the top by the plane wave with normal incidence along $x$ axis and polarization along $y$ axis. 
Absorption $Q_{abs}$ has been calculated using a set of monitors surrounding NP. 
Periodic boundary conditions have been applied at the lateral boundaries of the simulation box, while perfectly matched layer (PML) boundary conditions were used on the remaining top and bottom sides.
An adaptive mesh has been used to reproduce accurately the nanosphere shape. 

\begin{figure}
\centering
\includegraphics[width=120mm]{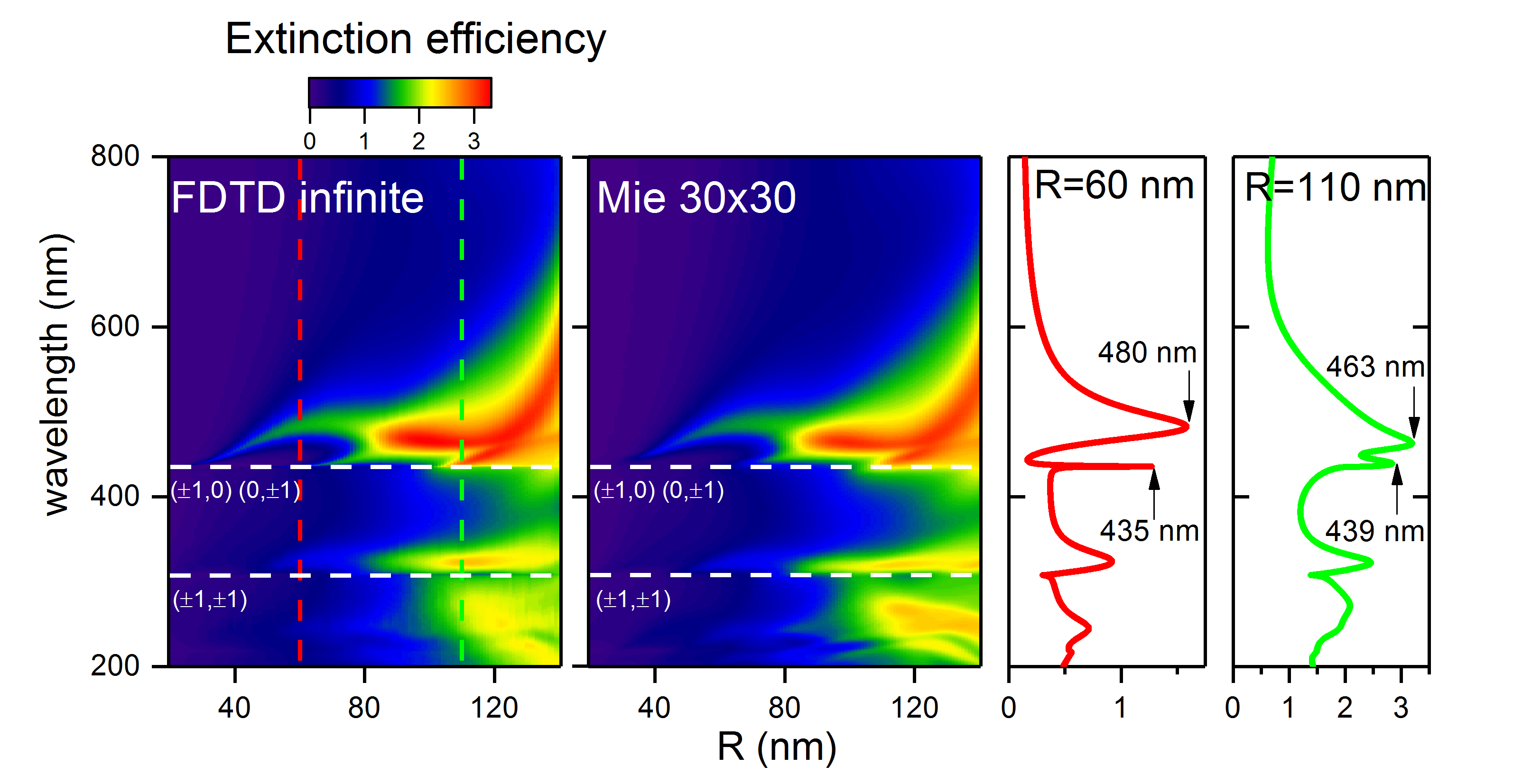}
\caption{Extinction spectra of the infinite (left) and $30 \times 30$ (right) arrays of Al NPs with $h=290$~nm for different values of $R$ calculated by FDTD and generalized Mie theory, respectively. White dashed line represent the different orders of the Rayleigh anomalies and white dots line represent the position of corresponding mode resonance.}
\label{fig:total_merge}
\end{figure}

\subsection{Generalized Mie Theory}

The generalized Mie theory~\cite{Mie1908} allows to calculate the extinction, scattering, and absorption spectra of spherical particles system. The incident and scattering coefficients are given by the equations
\begin{equation}
\begin{split}
    a^j_{mn}=\frac{\mu^j(m^j)^2j_n(m^jx^j) \left(x^j j_n(x^j) \right)^\prime-\mu^j j_n(x^j) \left(m^j x^j j_n(m^j x^j) \right)^\prime}{\mu^j(m^j)^2j_n(m^jx^j) \left(x^j h^{(1)}_n(x^j) \right)^\prime - \mu^j h^{(1)}_n(x^j) \left(m^j x^j j_n(m^j x^j) \right)^\prime}p^j_{mn},\\
    b^j_{mn}=\frac{\mu^j j_n(m^jx^j) \left(x^j j_n(x^j) \right)^\prime-\mu^j j_n(x^j) \left(m^j x^j j_n(m^j x^j) \right)^\prime}{\mu^j j_n(m^jx^j) \left(x^j h^{(1)}_n(x^j) \right)^\prime - \mu^j h^{(1)}_n(x^j) \left(m^j x^j j_n(m^j x^j) \right)^\prime}q^j_{mn}.
\end{split}
\label{mie}
\end{equation}
Here the superscript $j$ indicates the particle number.
The field at the $j$-th particle is the sum of the external field  and fields scattered by the other particles in the system:

\begin{equation}
\begin{split}
p^j_{mn}=p^{j,j}_{mn}-\sum\limits_{l = 1\hfill\atop
l \ne j\hfill}^N\sum_{\nu=1}^\infty\sum_{\mu=-\nu}^\nu \left(a^l_{\mu\nu} A^{\mu\nu}_{mn}(l,j)+b^l_{\mu\nu} B^{\mu\nu}_{mn}(l,j)\right),\\
q^j_{mn}=q^{j,j}_{mn}-\sum\limits_{l = 1\hfill\atop
l \ne j\hfill}^N\sum_{\nu=1}^\infty\sum_{\mu=-\nu}^\nu \left(a^l_{\mu\nu} B^{\mu\nu}_{mn}(l,j)+b^l_{\mu\nu} A^{\mu\nu}_{mn}(l,j)\right).
\end{split}
\label{gmm}
\end{equation}
Here $p^{j,j}_{mn}$, $q^{j,j}_{mn}$ are external field decomposition coefficients, $A^{\mu\nu}_{mn}(l,j)$, $B^{\mu\nu}_{mn}(l,j)$ are the translation coefficient~\cite{Xu1995}.
Equations~\eqref{mie},~\eqref{gmm} are the system of $n$ linear algebraic equations. The solution of this system provides one with coefficients $a^j_{mn}$, $b^j_{mn}$, $p^j_{mn}$, $q^j_{mn}$.

Extinction and scattering cross-sections of $n$-th mode reads as follows:
\begin{equation}
\begin{split}
(C_{ext})_n=\frac{4\pi}{k^2}n(n+1)(2n+2)\frac{(n-m)!}{(n+m)!}\sum_{j=1}^N\sum_{m=-n}^n Re(p^{j,j*}_{mn}a^j_{mn}+q^{j,j*}_{mn}b^j_{mn}),\\
(C_{sca})_n=\frac{4\pi}{k^2}n(n+1)(2n+2)\frac{(n-m)!}{(n+m)!}\sum_{j=1}^N\sum_{m=-n}^n \left( |a^j_{mn}|^2+|b^j_{mn}|^2 \right).
\end{split}
\end{equation}

\begin{figure}
\centering
\includegraphics[width=120mm]{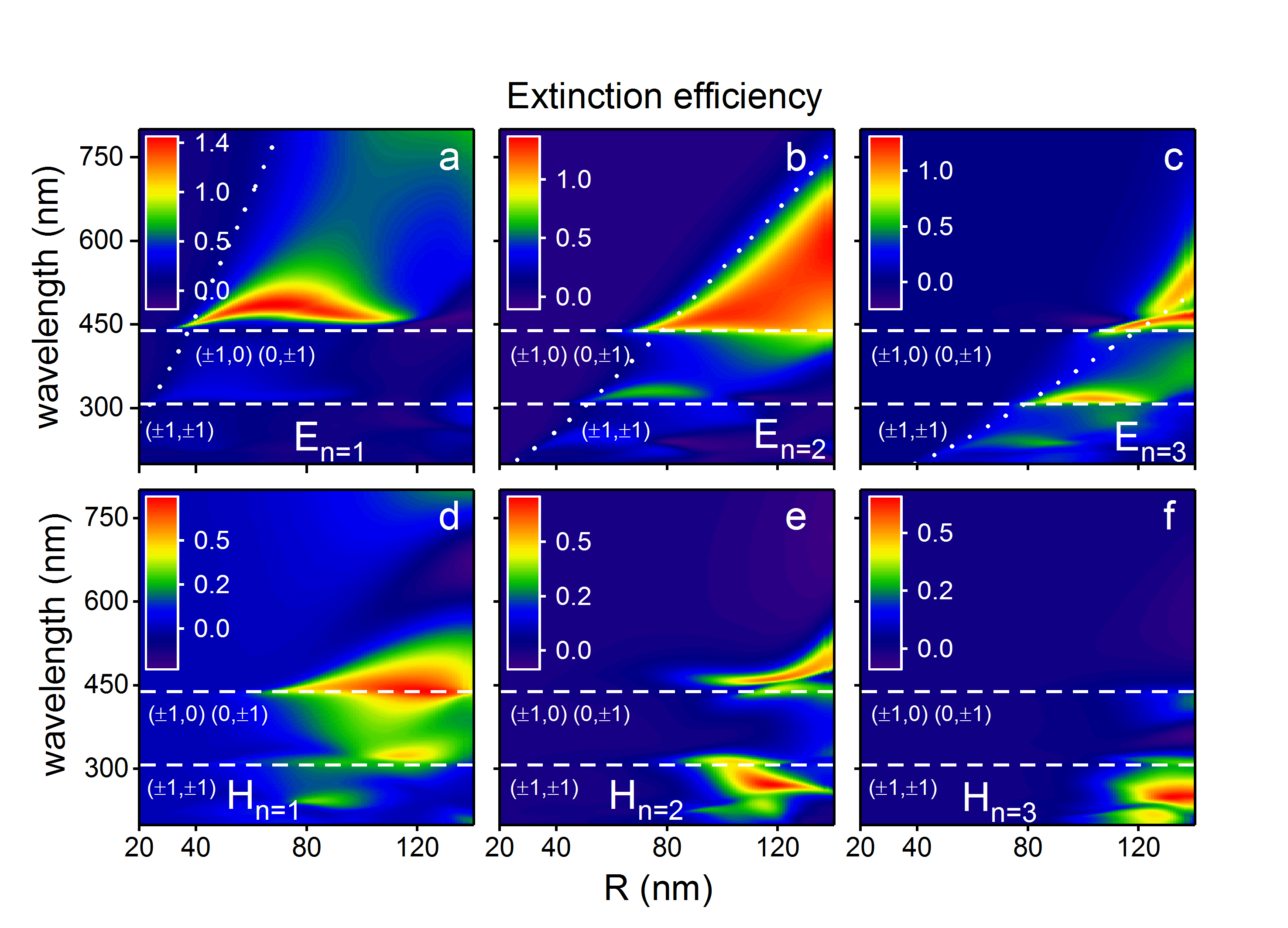}
\caption{Extinction efficiency decomposed in a set of spherical multipoles for square array consisting of Al NPs of different radius $R$ and wavelength of incidence light (a,b,c) are for the electrical field component while (d,e,f) are for magnetic field component. The array period is 290~nm. White dash lines represents Rayleigh anomalies and White dots lines correspond to the electric field resonance condition in Eq.(2) when denominator of $a^j_{mn}$ is equal to zero}
\label{fig:array_plot}
\end{figure}

\noindent
The total cross section is a sum cross-section of all modes:
\begin{equation}
C_{ext}=\sum_{n=1}^\infty(C_{ext})_n, \quad C_{sca}=\sum_{n=1}^\infty(C_{sca})_n.
\label{eq.Ext}
\end{equation}
The absorption cross-section is
\begin{equation}
C_{abs}=C_{ext}-C_{sca}.
\end{equation}

\noindent
The efficiency of extinction in this paper is understood as the ratio of the extinction coefficient to the area of the unit cell of the periodic structure:
\begin{equation}
Q_{ext}=\frac{C_{ext}}{h^2},
\label{efficiency}
\end{equation}
where $h$ is a lattice period.

\section{Results}
\subsection{Optical properties of a single aluminum sphere}

In Fig~\ref{fig:single_mult}a the extinction spectra of a single aluminum particle depending on its radius are shown.
It is seen that with increasing particle radius in the extinction spectrum the number of resonances also increases and their position is shifted to the long-wave region. 
In Fig~\ref{fig:single_mult} b-g the decomposed extinction spectra of the single nanoparticles for different radii are shown. As the one can see each mode is located in different  region in $(R,\lambda)$ space nevertheless they can overlap.

It should be noted that the increase in the radius of the particle leads to a sequential excitation of electric and magnetic modes. 
For example only electric dipole and quadrupole modes are excited in small particles. 
Octupole modes are excited in particles with radius greater than 50~nm.
It is important to note that the contribution of magnetic modes to extinction is much lower than the contribution of electric modes. 
In addition, these modes excites in particles with radius more than 80~nm.
At the same time, as the order of multipole decomposition increases, both the electric and magnetic modes are excited at shorter wavelengths.

\begin{figure}
\centering
\includegraphics[width=120mm]{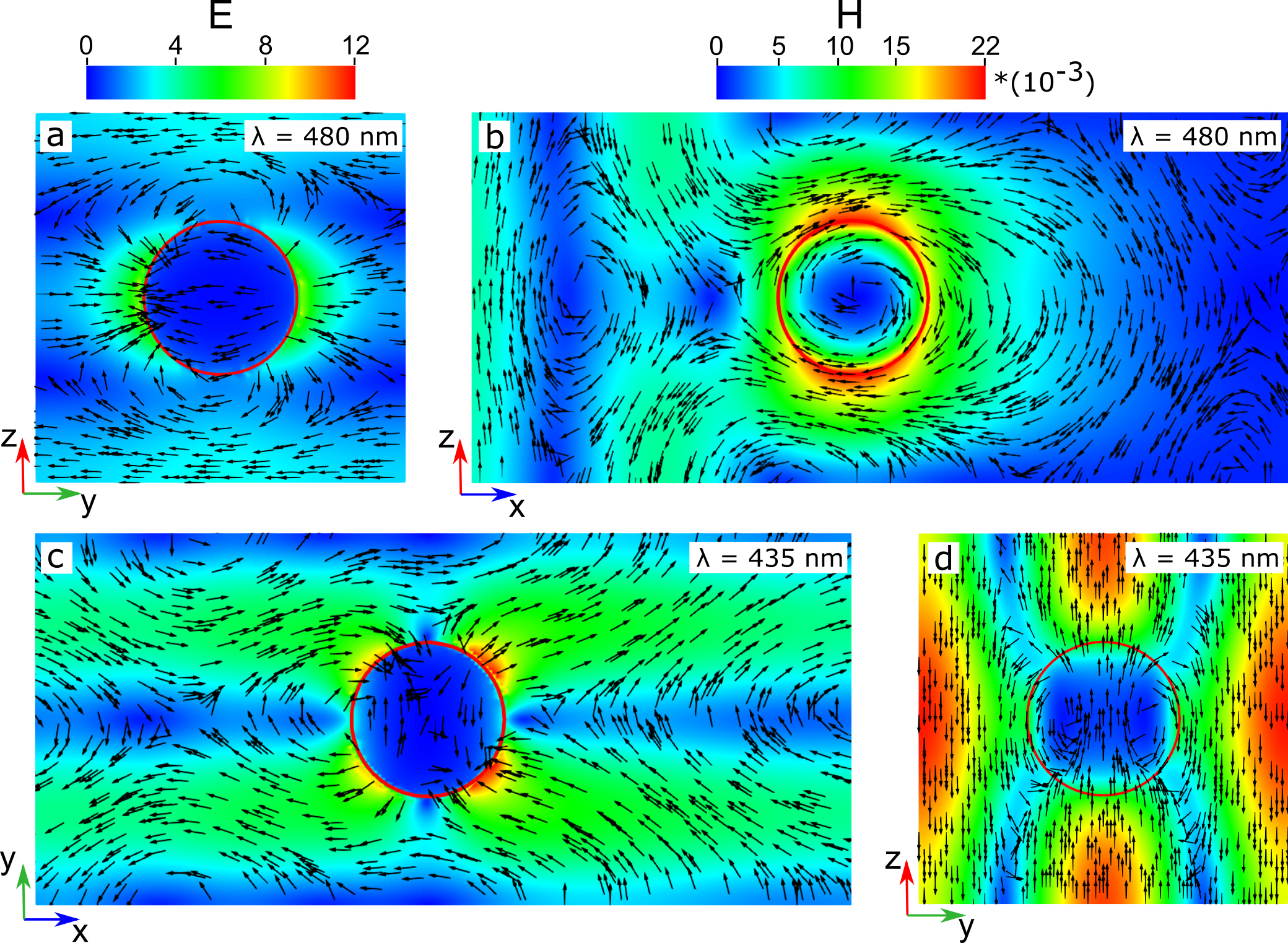}
\caption{The configuration of electric (a,c) and magnetic (b,d) fields at 435 and 480~nm, respectively for $R=60$~nm and $h=290$~nm.}
\label{fig:field}
\end{figure}

\subsection{Optical properties of a square array of aluminum spheres}
Using of aluminum nanospheres as structural elements of two-dimensional periodic arrays leads to a significant change in the extinction spectra.
This is due to hybridization of the single particle modes with the collective modes of the two-dimensional lattice.
Extinction spectra of the investigated periodic structures at different radii of the spherical particles and the periods of the lattice $h$ is presented in figure~\ref{fig:ot_p}.
It can be seen from these figures that the strong coupling between LSPR and Wood-Rayleigh anomalies~\cite{Wood1902,Rayleigh1907} leads to the emergence of high-quality collective resonances~\cite{Ross2016,Kravets2018} with spectral positions close to Wood-Rayleigh anomalies. 
Position of these anomalies for normal incidence of light for square arrays can be found with the following equation: 
\begin{equation}
\label{eq:Eq1}
\lambda_{s,q} = h \sqrt{ \frac{\varepsilon_{m}}{s^2 + q^2} }, 
\end{equation}
\noindent
where $s$, $q$ are integers which represent the order of the phase difference in $x$ and $y$ directions.
It should be noted that Eq.~(\ref{eq:Eq1}) describes condition of constructive interference for particles within the $YOZ$ plane~\cite{Bonod2016}. 
Here and after $\lambda$ is the vacuum wavelength.
Different orders of Wood-Rayleigh anomalies with $(0, \pm 1)$, $(\pm 1,0)$ and $(\pm 1, \pm 1)$ are shown in Fig.~\ref{fig:ot_p}.
Note, SLR in such structures can be tuned across the whole visible spectrum.

\begin{figure}
\centering
\includegraphics[width=120mm]{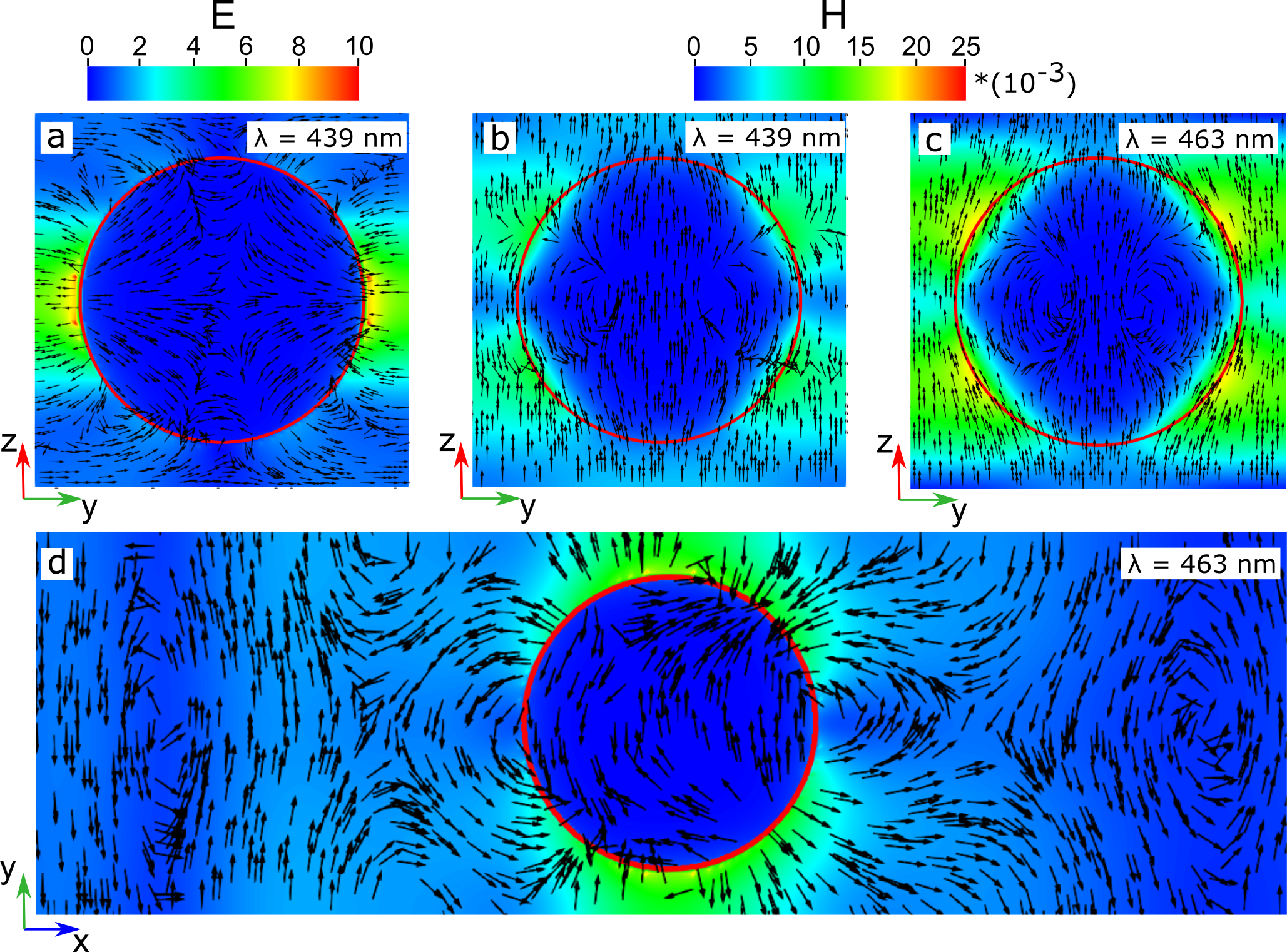}
\caption{The configuration of electric (a,d) and magnetic (b,c) fields at 439 and 463~nm, respectively for $R=110$~nm and $h=290$~nm.}
\label{fig:field2}
\end{figure}

\subsection{Finite size effects}
Let us turn to the dependence of extinction spectra on the radius of the particles.  In Fig.~\ref{fig:total_merge} the extinction spectra of an infinite lattice with a period of $h=290$~nm from aluminum spheres calculated by the FDTD method are presented.
To analyze the complex spectral pattern, we will use the generalized Mie theory, which will allow us to obtain the decomposition of extinction spectra by modes of spherical particles.

It should be noted that the generalized Mie theory allows us to calculate the optical properties of structures consisting of a finite number of particles. 
In this case, the extinction spectra of the two-dimensional lattice can be distorted by finite size effects (see \cite{Zakomirnyi2019}). 
Therefore, we have carried out comparative calculations of extinction spectra of an infinite periodic lattice and a lattice consisting of 30*30 particles. 
The calculation results depicted in Fig.~\ref{fig:total_merge}.

The figure shows that the use of an array of particles 30*30 with a high degree accurately reproduces the spectral characteristics of the infinite lattice. 
It is important to note that good agreement is observed as for the wavelengths of the modes of the two-dimensional lattice as their amplitudes. 
Thus, we can use generalized Mie theory to decompose the modes of a two-dimensional lattice over spherical multipoles with minimal impact of the finite size effects on them. 
The results of the expansions are shown in figure \ref{fig:array_plot}.

White dots lines correspond to the electric field resonance condition in Eq.\eqref{mie} when denominator of $a^j_{mn}$ is equal to zero. It should be noted that this equation is performed only for complex frequency. In this case the real part of frequency defines the position of the resonance while the imaginary part defines its line width. 
The spectral decomposition pattern is the result of the interaction of local resonances of single particles with  Rayleigh anomalies of the periodic array. 
For example the electric field mode $E_{n=1}$ is excited in the particles with the radius greater than 40~nm, while $E_{n=2}$ is excited in the particles larger than 80~nm. 
It is important to note that the excitation of $E_{n=3}$ is possible only in particles with a radius of about 120~nm.

An interesting fact is that the resonances of a single particle hybridize with higher-order Rayleigh anomalies, which leads to a local increase in the extinction efficiency near the wavelengths corresponding to the position of Rayleigh anomalies. 
A similar pattern is observed for the modes $E_{n=2}$ and $E_{n=3}$, while for the mode $E_{n=3}$ there is no increase in the extinction efficiency in the vicinity of the second Rayleigh anomaly with a wavelength of 300~nm.
This behavior can be explained by the extinction efficiencies of single particle  (Fig.~\ref{fig:single_mult}).
So for the $E_{n=1}$ mode, the Rayleigh anomaly crosses only the edge of the resonance region, while for $E_{n=2}$ and $E_{n=3}$ it crosses the region with high extinction efficiency.

\section{Resonant field configuration}
Now let us turn to the field configurations that correspond to the extinction maxima. 
For $R=60$~nm the maximum extinction achieved at the wavelength 435~nm and 480~nm while for $R=110$~nm at 439~nm and 463~nm~(see Fig.~\ref{fig:total_merge}). 
The corresponding field profiles for individual particle in the lattice are shown in Fig.~\ref{fig:field} and Fig.~\ref{fig:field2}.

As can be seen that at a wavelength of 480~nm the configuration of the electric and magnetic fields corresponds to the electric dipole.
While at a wavelength of 435~nm the field configuration resemble the combination of an electric quadrupole and a magnetic dipole. 
It is important to note that the electric dipole is excited in the plane of incident wave, and the quadrupole is in the plane of light polarization. 
A more complex configuration of the electric field is observed at a wavelength of 439~nm. 
 It is seen from figure Fig.~\ref{fig:field2} that in this case the third order mode is formed and it has three pairs of antisymmetric poles while modes of the second order are formed at the wavelength of 463~nm  Fig.~\ref{fig:field2}d. 
Note that both at the wavelength of 439~nm and at the wavelength of 463~nm, the configuration of magnetic fields resemble the magnetic dipole.

For particles with a radius of 110~nm and a wavelength of incident light of 448~nm, magnetic modes of only the first order can be excited in such media, while a change in the wavelength up to 458~nm leads to the excitation of magnetic modes of the second order.
The obtained field distributions are in full agreement with the results of the decomposition of the extinction spectra obtained by the generalized Mie theory. 
Thus, the variation of the lattice parameters and the wavelength of the incident light makes it possible to effectively control the modes of two-dimensional periodic lattices.

\section{Conclusion}
To conclude, extinction spectra of the single aluminium nanoparticles and NPs arrays depending on particle radius were obtain. 
It is shown that for single nanoparticles  the increase of the radius leads to a sequential excitation and subsequent decay of electric and magnetic modes. 
It should be noted that the modes of a single particle do not interact with each other.
However, a different picture is observed for the NPs array. 
In this case the different orders of particle modes interact with each other and with Rayleigh anomalies that leads to the formation of hybrid modes.
The spectral manifestations of hybrid modes are the formation of narrow lines in the extinction spectra and the corresponding hybrid configurations of the electromagnetic field.
The feature here is that the each mode of both electric and magnetic fields interact with Rayleigh anomalies of different orders. 
As the result we observe increased extinction efficiency at the vicinity of  Rayleigh anomalies. This effect can be used to tailor the spectral position of extinction maximum.    
Thus, understanding of modes coupling in such system paves a way for more efficient control of its optical response for photonics applications which is not easy to achieve with other alternative strategies.

\section*{Funding}
The reported study was funded by the Russian Foundation for Basic Research, Government of Krasnoyarsk Territory, Krasnoyarsk Regional Fund of Science (Grant No.18-42-240013); A.E. thanks the grant of the President of Russian Federation (agreement 075-15-2019-676).

\section{Acknowledgement}
We thank Ilia Rasskazov and Polina Semina for fruitful and valuable discussions.

\bibliographystyle{prsty}
\bibliography{references_Ilia,biblio}
\end{document}